\documentclass{webofc}
\usepackage[varg]{txfonts}   
\begin{document}
\title{Studying exotic hadrons in high energy nuclear collisions}
%
%

\author{\firstname{Xingyu} \lastname{Guo}\inst{1,2}\fnsep\thanks{\email{guoxy@m.scnu.edu.cn}} \and
        \firstname{Jinfeng} \lastname{Liao}\inst{4}\fnsep\thanks{\email{liaoji@indiana.edu}} \and
        \firstname{Hongxi} \lastname{Xing}\inst{1,2,3}\fnsep\thanks{\email{hxing@m.scnu.edu.cn}}
}

\institute{ 
Key Laboratory of Atomic and Subatomic Structure and Quantum Control (MOE), Guangdong Basic Research Center of Excellence for Structure and Fundamental Interactions of Matter, Institute of Quantum Matter, South China Normal University, Guangzhou 510006, China
\and
          Guangdong-Hong Kong Joint Laboratory of Quantum
Matter, Guangdong Provincial Key Laboratory of Nuclear Science, 
Southern Nuclear Science Computing Center, South China Normal University, Guangzhou 510006, China
\and 
Southern Center for Nuclear-Science Theory (SCNT), Institute of Modern Physics, 
Chinese Academy of Sciences, Huizhou 516000, China
\and 
          Physics Department and Center for Exploration of Energy and Matter,
Indiana University, 2401 N Milo B. Sampson Lane, Bloomington, IN 47408, USA
          }

\abstract{%
  Studies of exotic hadrons such as the $\chi_{c1} (3872)$ state provide crucial insights into the fundamental force governing the strong interaction dynamics, with an emerging new frontier to investigate their production in high energy nuclear collisions where a partonic medium is present. This contribution discusses the production mechanisms of exotic hadrons in such collisions and analyzes novel effects from the partonic medium, demonstrating the potential to use heavy ion measurements for deciphering their internal structure and understanding their in-medium evolutions.     
}
\maketitle
\vspace{-0.5cm}

\section{Introduction}
\label{intro}

Understanding the nature of exotic multiquark candidates like  the $\chi_{c1} (3872)$  is an outstanding challenge in strong interaction physics.  While properties of  $\chi_{c1} (3872)$ have been studied at multiple colliders, physicists remain puzzled by the nature of this particle despite 20 years past its initial discovery, with multiple interpretations being proposed for its internal structures, such as a   large-size hadronic molecule versus  a compact-size tetraquark state. A new avenue of investigating exotic hadrons has recently emerged and rapidly developing, namely to study their formation in high energy hadron and nuclear collisions where a partonic medium is present. In such collisions, a fireball with 
  many light flavor quarks/antiquarks (on the order of hundreds to thousands depending on colliding systems) 
is created together with a considerable number of charm quarks/antiquarks. This provides an ideal environment for creating heavy flavor exotic states and probing their properties, as demonstrated in the latest theoretical works~\cite{ExHIC:2010gcb,ExHIC:2017smd,Zhang:2020dwn,Esposito:2020ywk,Braaten:2020iqw,Wu:2020zbx,Chen:2021akx,Hu:2021gdg,Guo:2023dwf}.    
 Most importantly, experimental measurements of $\chi_{c1} (3872)$ production in these collisions have started to arrive in the last few years, including LHCb data from high multiplicity proton-proton (pp) collisions~\cite{LHCb:2020sey} and proton-lead (pPb) collisions~\cite{LHCb:2022ixc} as well as CMS data from lead-lead (PbPb) collisions~\cite{CMS:2021znk} at the Large Hadron Collider (LHC). Already this first batch of empirical information shows an unusual pattern of the partonic medium's influence on the $\chi_{c1} (3872)$ yield with respect to the yield of another particle called  $\psi(2S)$ which is a normal hadronic state serving as a benchmark for comparison. In this contribution, we discuss the production mechanisms of exotic hadrons in high energy nuclear collisions and analyze novel effects from the partonic medium, demonstrating the potential to use heavy ion measurements for deciphering their internal structure and understanding their in-medium evolutions.

\section{Volume effect and the intrinsic size of $\chi_{c1} (3872)$} 
\label{sec-1}

The production of $\chi_{c1} (3872)$ in the soft region with relatively low transverse momentum is presumably dominated by coalescence mechanism. Depending on whether it is a compact tetraquark or a hadronic molecule, it can be formed at freeze-out in different ways. In the former case, one could imagine the formation of a diquark ($cq$) cluster and an anti-diquark ($\bar{c}\bar{q}$)  cluster within a small volume (as constrained by QCD color confinement) which then coalesce into a compact $\chi_{c1} (3872)$. In the latter case, two D mesons ($c\bar{q}$ and $\bar{c}q$) need to be first formed at large relative distance and then coalesce into a molecular state.

\begin{figure}[hbt!]
\centering
\sidecaption
\includegraphics[width=6cm,clip]{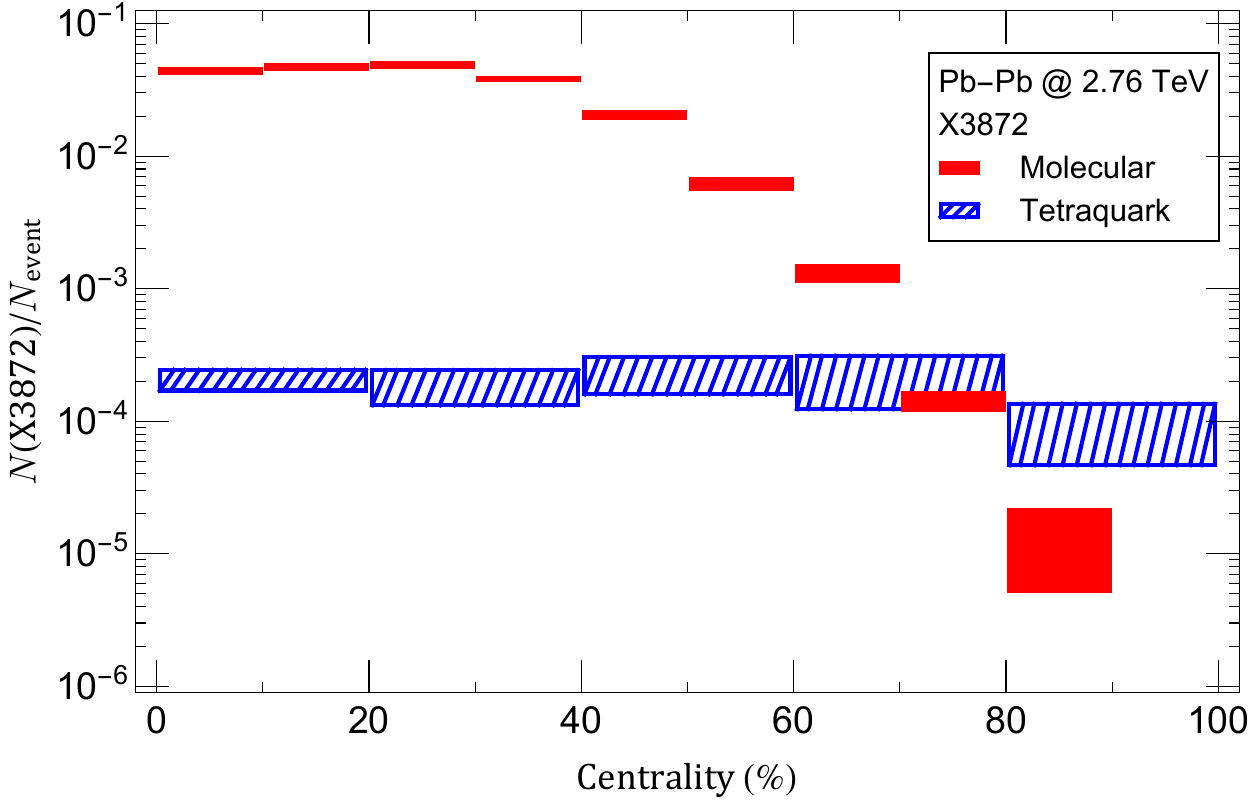}
\caption{The centrality dependence of the $\chi_{c1} (3872)$ in Pb-Pb collisions at $\sqrt{s}=2.76~\mathrm{TeV}$ for hadronic molecular configuration (red solid boxes) and tetraquark configuration (blue shaded boxes). See  \cite{Zhang:2020dwn} for details. }
\vspace{-0.2cm}
\label{fig-1}       
\end{figure}

To examine the impact of its internal structures on the production, 
a quantitative calculation was performed in \cite{Zhang:2020dwn}  by adopting a multiphase transport model (AMPT)  for describing bulk evolutions  and implementing appropriate   production mechanism of either molecule or tetraquark scenario. The results suggest a significantly larger yield of the $\chi_{c1} (3872)$ as hadronic molecules, by two orders of magnitude, than the compact tetraquark case. Clearly the interplay between the fireball volume and the  $\chi_{c1} (3872)$ intrinsic size  plays a significant role. This is further demonstrated by the centrality dependence of the $\chi_{c1} (3872)$ yield shown in Fig. \ref{fig-1}. Going from central to peripheral collisions, one observes a strong decrease  for the molecular scenario while a  rather mild change for the tetraquark scenario. As a baseline of expectation, the available number of $c$ and $\bar{c}$ quarks would gradually decrease with increasing centrality class, with the fireball spatial volume and evolution time also decreasing. The sharp decrease of molecular state production toward very peripheral collision is due to the shrinking volume available for accommodating the large-size hadronic molecule.  The relatively flat dependence of the tetraquark case is due to two compensating factors: decreasing numbers of $c/\bar{c}$ quarks while increasing chances of small spatial separation between (anti-)diquarks due to shrinking fireball volume.  Similar results were also obtained in \cite{Hu:2021gdg} for the production of another exotic state $T_{cc}^+$. In addition to centrality, one would also expect a similar 
 system-size dependence of $\chi_{c1} (3872)$ production across different colliding systems from OO and ArAr to  XeXe and PbPb. 
 Such volume effect provides a unique opportunity for determining the intrinsic size of $\chi_{c1} (3872)$ and 
 distinguishing hadronic molecule versus compact tetraquark scenarios, which would help  address a long-standing hadron physics challenge with heavy ion  measurements.

\section{Medium-assisted production of $\chi_{c1} (3872)$}
\label{sec-2}

\begin{figure}[hbt!]
\centering 
\sidecaption
\includegraphics[width=6.cm,clip]{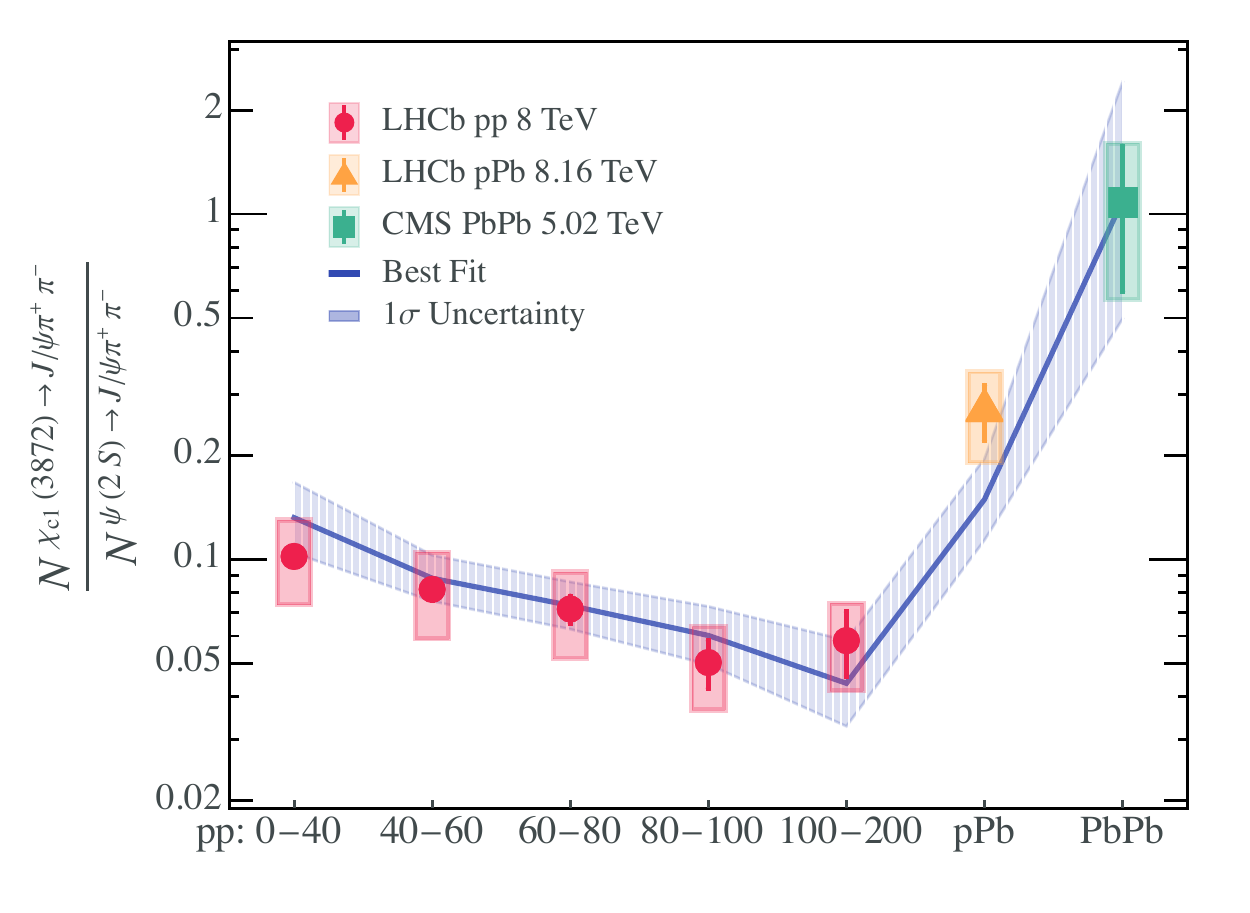}
\caption{A comparison of the $\chi_{c1} (3872)$ yield relative to $\psi(2S)$ between model simulation results (blue curve) and experimental data from LHCb pp collisions at $\sqrt{s_{NN}}=8$ TeV (red circle),   LHCb preliminary pPb collisions at $\sqrt{s_{NN}}=8.16$ TeV (orange triangle) and  CMS PbPb collisions at $\sqrt{s_{NN}}=5.02$ TeV (green box)~\cite{CMS:2021znk,LHCb:2020sey,LHCb:2022ixc}. 		
The model parameters are determined from the global fitting analysis with the blue band showing the $1\sigma$ level uncertainty. }
\vspace{-0.3cm}
\label{fig-2}       
\end{figure}

In the hard region with relatively high transverse momentum, production of $\chi_{c1} (3872)$ should dominantly come from virtual  $c\bar{c}$ pairs that are generated in the initial hard scatterings and subsequently traverse the partonic medium before turning into final hadronic states. 
We note this is the more relevant situation for the recent experimental measurements, including those from LHCb pp (at $\sqrt{s_{NN}}=8$ TeV)  and preliminary pPb (at $\sqrt{s_{NN}}=8.16$TeV) as well as the CMS PbPb (at $\sqrt{s_{NN}}=5.02$ TeV) analyses. These data show an intriguing evolution pattern of the $\chi_{c1} (3872)$-to-$\psi(2S)$ yield ratio from proton-proton collisions with increasing multiplicities toward proton-lead and lead-lead collisions, see Fig.~\ref{fig-2}. 

In order to explain the observed pattern, one needs to understand the medium attentuation effects on their production. 
The first important effect is the medium absorption. Random collisions with quarks and gluons from the medium result in the dissociation of the correlated co-moving $c\bar{c}$ pair, which leads to a suppression on the final hadron yield. 
In addition to this, we propose a novel mechanism of medium-assisted enhancement for the $\chi_{c1} (3872)$ production. This is because the formation of $\chi_{c1} (3872)$ requires two light quarks/antiquarks in addition to the $c\bar{c}$. Scatterings with the partonic medium, which serves as a reservoir of numerous light quarks/antiquarks,  could lead to ``picking up'' of light quarks/antiquarks which then co-move with the $c\bar{c}$ pair. This increases  the probability to form the $\chi_{c1} (3872)$ state in the end. Such medium-assisted enhancement competes with the  conventional absorption-induced suppression and results in a nontrivial  trend from small to large colliding systems.

To quantitatively examine these medium effects, we implement the above model within event-by-event hydrodynamic  simulations and compute the $\chi_{c1} (3872)$-to-$\psi(2S)$ yield ratio. See \cite{Guo:2023dwf} for details.  Two key parameters $\alpha'$ and $\beta'$ are introduced in the model to quantify the strength of in-medium absorption and enhancement, respectively. Their values are determined from a global fitting analysis with the available LHCb and CMS data: see Fig.~\ref{fig-2}.  As one can see, our model with just two parameters   provides a first quantitative description of these data   from small to large systems. The non-monotonic behavior is a consequence of the competition   between suppression and enhancement, with the former dominating for relatively low medium density and/or small medium size while the latter dominating for relatively high medium density and large medium size.

\begin{figure*}
\centering
\includegraphics[width=4.5cm,clip]{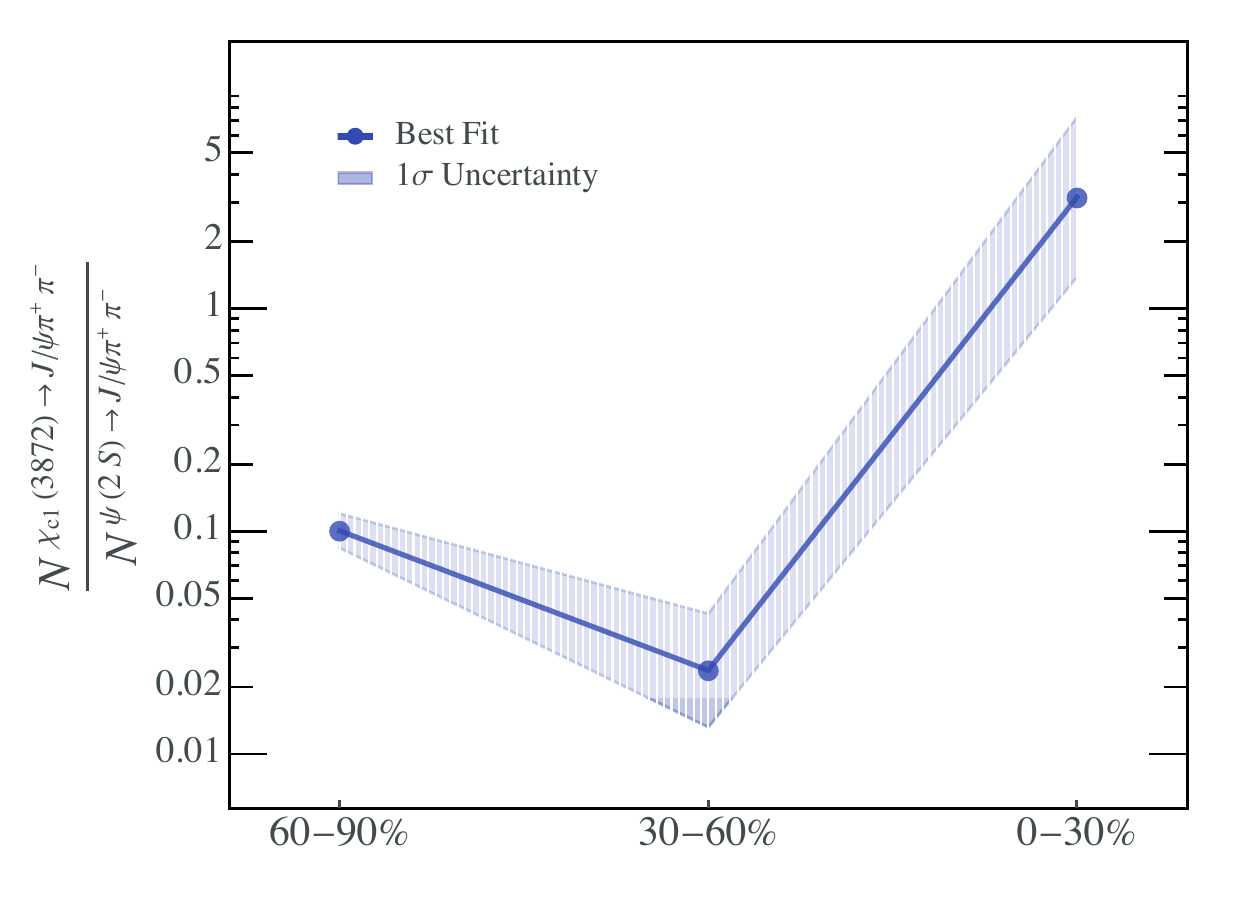} \hspace{0.5cm}
\includegraphics[width=4.5cm,clip]{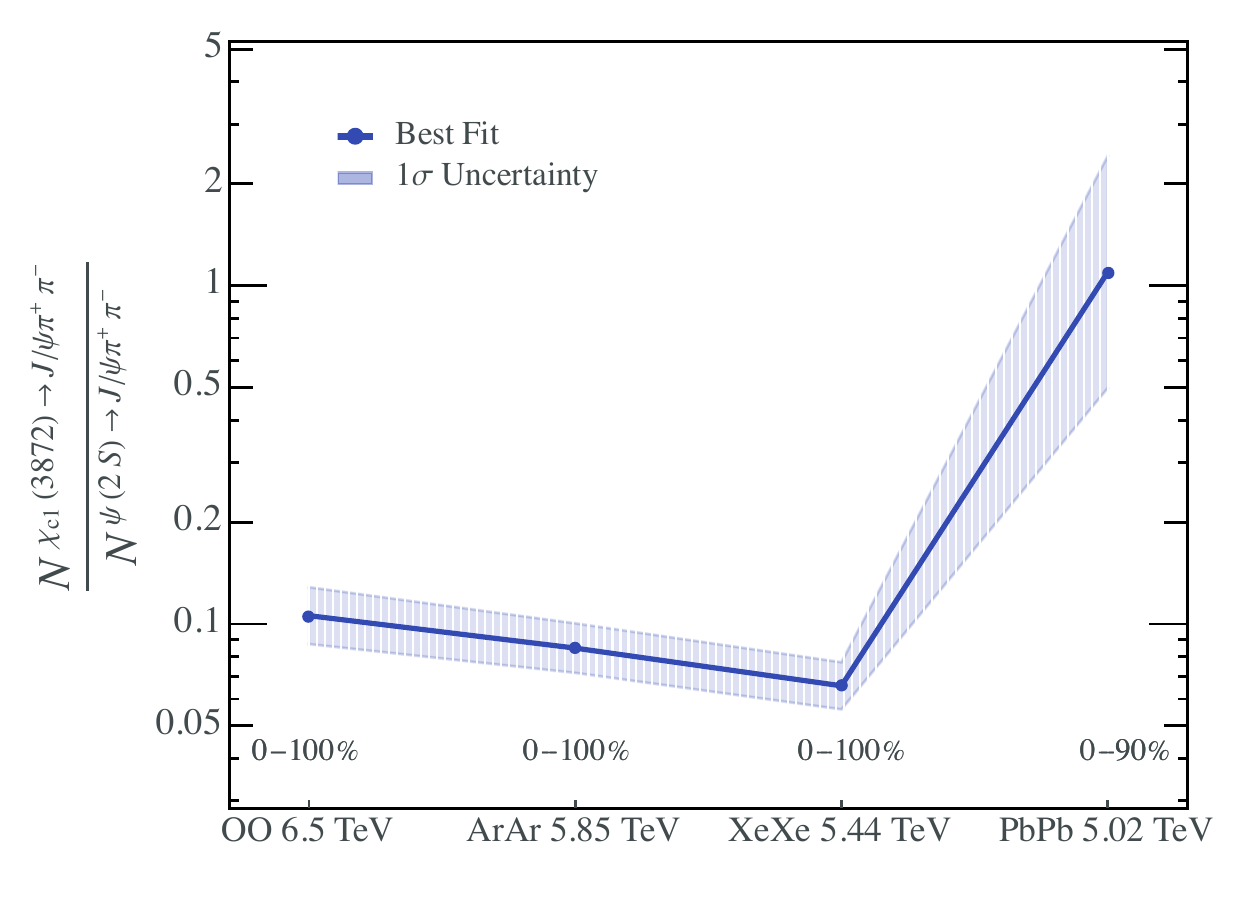}
\caption{ Predictions for centrality dependence in PbPb collisions (left) and system size dependence (right) of the $\chi_{c1} (3872)$ yield relative to $\psi(2S)$. See \cite{Guo:2023dwf} for details. }
\vspace{-0.5cm}
\label{fig-3}       
\end{figure*}

A natural next step would be testing our model predictions for which experimental data are not yet available and can serve as a future validation. To do that, we compute the centrality dependence of the $\chi_{c1} (3872)$-to-$\psi(2S)$ yield ratio   in PbPb collisions as well as the system size dependence from OO and ArAr to XeXe and PbPb collisions, with the results shown in Fig.~\ref{fig-3}. In both cases, a non-monotonic behavior emerges as the imprint of the competition between enhancement and suppression, which can be readily tested by future data.


\section{Summary}

To summarize, the partonic medium created in high energy nuclear collisions provides a unique environment for producing heavy exotic states  such as the $\chi_{c1} (3872)$ and studying their structures. As demonstrated by recent calculations, the medium size provides a natural and valuable ``meter stick'' for calibrating the intrinsic size of $\chi_{c1} (3872)$ production in the soft region due to volume effect in their coalescence formation.  For the production of $\chi_{c1} (3872)$ in the hard region, a phenomenological model for the partonic medium attenuation effects is presented for explaining the experimental data from LHCb and CMS. In particular, a novel mechanism of medium-assisted enhancement effect is proposed for the $\chi_{c1} (3872)$ production, which leads to a competition with the more conventional absorption-induced suppression effect and becomes more dominant for higher parton densities and larger medium size. Simulations from this model offers the first quantitative description of currently available measurements from LHCb and CMS. Further predictions are made for the centrality dependence of the $\chi_{c1} (3872)$-to-$\psi(2S)$ yield ratio in PbPb collisions as well as for its system size dependence from OO and ArAr to XeXe and PbPb collisions, which can be tested by future high precision measurements.



{\it Acknowledgments.}
This research was supported in part by the National Natural Science Foundation of China (NSFC) under Grants No.~12035007, No.~12022512 and No.~11905066, by Guangdong Major Project of Basic and Applied Basic Research No.~2020B0301030008 as well as by the U.S. NSF (Grant No.~PHY-2209183).


%
%
%
%
%

\end{document}